\documentclass[twocolumn,preprintnumbers,showpacs]{revtex4-1}
\usepackage{amsmath}
\usepackage{amssymb}
\usepackage{graphicx}
\usepackage{color}

\newcommand{\BAN}{\ensuremath{B_{1g}\,}}
\newcommand{\AO}{\ensuremath{A_{1g}\,}}
\newcommand{\AOO}{\ensuremath{A'_{1g}\,}}
\newcommand{\AOT}{\ensuremath{A''_{1g}\,}}
\newcommand{\AT}{\ensuremath{A_{2g}\,}}
\newcommand{\BN}{\ensuremath{B_{2g}\,}}

\begin{document}

\title{Three energy scales in the superconducting state of hole-doped cuprates detected by electronic Raman scattering}

\author{S. Benhabib$^1$,Y. Gallais$^1$, M. Cazayous$^1$,  M.-A. M\'easson$^1$, R. D. Zhong$^2$, J. Schneeloch$^2$, A. Forget$^3$, G. D. Gu$^2$, D. Colson$^3$ and A. Sacuto$^1$ }

\affiliation{$^1$ Laboratoire Mat\'eriaux et Ph\'enom$\grave{e}$nes Quantiques (UMR 7162 CNRS),
Universit\'e Paris Diderot-Paris 7, Bat. Condorcet, 75205 Paris Cedex 13, France,\\
$^2$ Matter Physics and Materials Science, Brookhaven National Laboratory (BNL), Upton, NY 11973, USA,\\
$^3$Service de Physique de l'Etat Condens\'{e}, CEA-Saclay, 91191 Gif-sur-Yvette, France}

\date{\today}

\begin{abstract}

We explored by electronic Raman scattering the superconducting state of Bi$_2$Sr$_2$CaCu$_2$O$_{8+\delta}$ (Bi-2212) crystal by performing a fine tuned doping study. We found three distinct energy scales in \AO,\BAN and \BN symmetries which show three distinct doping dependencies. Above p=0.22 the three energies merge, below p=0.12, the \AO scale is no more detectable while the \BAN and \BN scales become constant in energy. In between, the \AO and \BAN scales increase monotonically with under-doping  while the \BN one exhibits a maximum at p=0.16. The three superconducting energy scales appear to be an universal feature of hole-doped cuprates. We propose that the non trivial doping dependencies of the three scales originate from the Fermi surface changes and reveal competing orders inside the superconducting dome.

\end{abstract}


\maketitle


\section*{Introduction}

Conventional superconductors are characterized by a single energy scale, the superconducting gap, which is proportional to the critical temperature $T_{c}$~\cite{Bardeen57}. The existence of more than one energy scale in the superconducting state of hole-doped cuprates has stirred many debates since several years~\cite{Tallon1994,Deutscher1999,Opel2000,Wen2005,LeTacon06,Tanaka2006,Valla2006,Boyer2007} and has raised the question of the existence of more than one gap in the superconducting state of cuprates~\cite{Tallon01, Millis2006,Huefner08,Guyard2008a,Kondo2009}. To move toward an understanding of high-temperature superconductivity, one of the most challenging issue is the identification of the energy scales associated with the onset of coherent excitations in the superconducting state. The electronic Raman spectroscopy is an efficient probe for this task. Depending on the symmetries (\AO, \BAN or \BN) related to the quasi-tetragonal structure of the cuprates, the energy scales can be explored in different regions of the Brillouin zone by electronic Raman scattering. Usually, the \BAN and \BN  energy scales are respectively assigned to the maximum amplitude of the $d-$wave $2\Delta$ pairing gap in the region near $(\pm\pi,0)$ and $(0,\pm\pi)$ (called the antinodal region) and the weaker amplitude in the region near $(\pm\pi/2,\pm\pi/2)$ (called the nodal region) ~\cite{Devereaux-RMP,Sacuto13}.
A priori these two scales have the same origin (the $2\Delta$ pairing gap) and have to follow each other. In fact, they exhibit distinct doping dependencies and are responsible for the two gaps issue in the superconducting state of hole-doped cuprates. Several distinct scenarios are still debated ~\cite{Chen1997,Tallon01,Wen2005,LeTacon06,Tanaka2006,Millis2006,Huefner08,Chubukov2008,Valenzuela2008,Blanc2010,LeBlanc2010,Vishik12,Sacuto13}. One of them has been to associate the nodal energy scale to the superconducting state while the anti-nodal one is associated with the pseudogap~\cite{Alloul89,Warren89}. Although the origin and significance of these two scales are not yet explained a common thread is emerging: at least two electronic orders compete inside the superconducting dome of hole-doped cuprates and make distinct the doping evolutions of \BAN and \BN energy scales.

Competing orders are the necessary ingredients to induce Fermi surface changes \cite{Abanov03,Norman2010}. 
In hole-doped cuprates angular photoemission spectroscopy (ARPES) revealed above $T_c$ a disconnected Fermi surface with "Fermi arcs" centered around the nodes in underdoped Bi-2212  compound \cite{Norman1998}. More recently, quantum oscillations measurements under high magnetic field and low temperature have shown that Fermi surface undergoes a reconstruction into pockets in underdoped YBa$_2$Cu$_3$O$_{7-\delta}$ (Y-123) compound \cite{DoironLeyraud07,Sebastian2012,Sebastian2014,DoironLeyraud2015}. 

On the other hand, there exists a third energy scale detected in \AO symmetry whose origin is still mysterious although it has been detected for a long time \cite{Cooper88,Staufer92,Sacuto97,Gallais04,LeTacon05b}. In Y-123 compound, the \AO energy peak was found to follow the inelastic neutron scattering resonance \cite{Rossat,Bourges,Hinkov04} with nickel and zinc substitutions \cite{SacutoSidis02,LeTacon06b}. In the very beginning of the cuprates Raman studies, the \AO peak was assigned to the $2\Delta$ pairing gap, but later considerations showed that long range Coulomb screening washes out the gap effects in the $A_{1g}$ geometry \cite{cardona97,Devereaux-RMP} and recent investigations advocate rather in favor of a collective mode \cite{Venturini2000,Devereaux-RMP,Montiel2015}. However, no thorough doping evolution of the \AO energy scale has been yet established.

In this study,  our purpose is to accurately determine the actual doping dependencies of these three energy scales on a large range of doping \textsl{p} to get a better understanding of their origins. In particular, we are interested in finding the specific range of doping levels for which the energy scales present drastic changes and how they can be connected to the doping evolution of the Fermi surface and competing orders mentioned above.

In order to reach this goal we have performed light polarized electronic Raman scattering on Bi-2212 single crystals which can be associated to the I/4mmm-$D^{17}_{4h}$ space group \cite{Note}. We have got the Raman spectra in \AO, \BAN and \BN symmetries. In particular special care has been devoted to extract the pure \AO spectrum from well controlled subtractions of the Raman spectra. Importantly, the doping level was solely controlled by oxygen insertion to avoid cationic substitutions which can drastically change the electronic Raman spectra \cite{Munnikes2011}. The large range of doping levels from p=0.06 to p=0.23 was successfully obtained from specific annealing treatments described in the experimental procedure. This allows us to follow simultaneously in an unique system the doping evolution of the \AO, \BAN and \BN energy scales and track them even for from very low and high doping levels.  

We show that the \AO, \BAN and \BN energy scales in Bi-2212 merge together above p=0.22 and decrease with doping. This corresponds to a huge enhancement of the antinodal Bogoliubov quasiparticles spectral weight related to a Lifshitz transition where the hole like Fermi surface transforms into an electron like. On the other hand, the difference between the \BAN and \BN energy scales is abnormally large below p=0.12 and the \BAN and \BN scales are almost constant with doping. This corresponds to a significant loss of the antinodal Bogoliubov quasiparticles spectral weight where charge ordering settled. Between these two doping levels the \AO and \BAN scales increase monotonically while the \BN scale is non monotonic and it is peaked at p=0.16 for which $T_{c}$ is maximum.

\section*{Details of the Experimental Procedure}

\subsection*{A. Raman Experimental Set Up }

Electronic Raman experiments have been carried out using a triple grating spectrometer (JY-T64000) equipped with a liquid-nitrogen-cooled CCD detector. Raman spectra above and below $T_c$ were obtained using an ARS closed-cycle He-cryostat. The laser excitation line used was the 532 nm of a diode pump solid state laser. The laser power at the entrance of the cryostat was maintained below $2~mW$ to avoid over heating of the crystal estimated to $3~K/mW$ at $10~K$. The \BAN+\AT and \BN+\AT symmetries have been obtained from cross linear polarizations at 45$^o$ from the Cu-O bond directions and along them respectively~\cite{Sacuto2011}. The change between these both symmetries was obtained by keeping fixed the orientations of the analyzor and the polarizor and by rotating the crystal with an Attocube piezo-driven rotator. We  got an accuracy on the crystallographic axes orientation with respect to the polarizors close to $2^o$. The \AO+\BN and \AO+\BAN symmetries were obtained from linear parallel polarizations at 45$^o$ from the Cu-O bond directions and along them respectively. In practice we measure the \BAN+\AT and \AO+\BN responses and then rotate the crystal to get the \BN+\AT and \AO+\BAN  ones. In the following we have made the assumption that the \AT electronic Raman scattering contribution is negligible i.e: the Raman vertices $\gamma_{xy}-\gamma_{yx}\approx0$. This is supported by the very weak Raman contribution at low energy (below 1000~$cm^{-1}$) in pure \AT extracted from a combination of linear and circular polarizations spectra~\cite{venturiniphd}.

All the spectra have been corrected for the Bose factor and the instrumental spectral response. They are thus proportional to the imaginary part of the Raman response function $\chi^{\prime \prime}_{\nu} (\omega,T)$ where $\nu$ refers to the vertex symmetry \AO, \BAN or \BN.

\subsection*{B. Crystal Growth and Characterization}

The Bi-2212 single crystals were grown by using a floating zone method. The optimal doped sample with $T_{c} = 90~K$ was grown
at a velocity of 0.2 mm per hour in air ~\cite{Wen_b}. In order to get overdoped samples down to $T_{c}=65~K$ , the as-grown single crystal was put into a high oxygen pressured cell between $1000$ and $2000$ bars and then was annealed from $350^{o}C$ to $500^{o}C$ during 3 days ~\cite{Mihaly}. The overdoped samples below $T_{c}=60~K$ was obtained from as-grown Bi-2212 single crystals put into a pressure cell (Autoclave France) with $100$ bars oxygen pressure and annealed from $9$ to $12$ days at $350~^{o}C$. Then the samples were rapidly cooled down to room temperature by maintaining a pressure of $100$ bars. In order to get the underdoped sample down to $T_{c}= 50~K$, the optimal doping crystal was annealed between $350~^{o}C$ and $550~^{o}C$ during 3 days under vacuum of $1.3~10^{-6}~mbar$.  The critical temperature $T_{c}$ for each crystal has been determined from magnetization susceptibility measurements at a $10$ Gauss field parallel to the c-axis of the crystal. More than 30 crystals have been measured among $60$ tested. The selected crystals exhibit a quality factor of $T_{c}/ \Delta T_{c}$ larger than $7$. $\Delta T_{c}$ is the full width of $T_{c}$ transition measured. A complementary estimate of $T_{c}$  was achieved from electronic Raman scattering measurements by defining the temperature from which the \BAN superconducting pair breaking peak collapses. 
The level of doping $p$ was defined from $T_c$ using Presland and Tallon's equation \cite{Presland}: $1-T_{c}/T_{c}^{max} = 82.6 (p-0.16)^{2}$. $T_{c}$ versus p is reported in fig.3 (a).

\section*{Experimental Results}

In figure 1 (a-d) are displayed the Raman responses of an over-doped (OD) Bi-2212 single crystal with a $T_{c}=60~K$. The (red/grey) and (black) curves were measured below and above $T_{c}$ in the \AO+\BAN, \AO+\BN, \BAN and \BN symmetries. The sharp peaks located at 127, 297, 327, 359 and 474~$cm^{-1}$ are phonon lines that will be discussed in a next article. A comparison between (c) and (d) panels shows that the \BAN electronic contribution is preponderant compared to the \BN one. It manifests itself as an intensive $2\Delta$ pair breaking whose maximum is peaked at 244~$cm^{-1}$ (see inset of panel (c)). The \BN electronic peak although much lower in intensity is centered around the same energy ($\approx254~cm^{-1}$). 

In \AO+\BAN symmetry (panel (a)),  the \BAN electronic contribution is dominant such that the \AO electronic peak can only be observed in the \AO+\BN response (panel (b)) since the \BN contribution is weak. The \AO electronic continuum (detectable in panel (b)) is maximum around 250~$cm^{-1}$ nearby the energies of the \BAN and \BN peaks.

\begin{figure}[bp]
\begin{center}
\includegraphics[width=9cm,height=10cm]{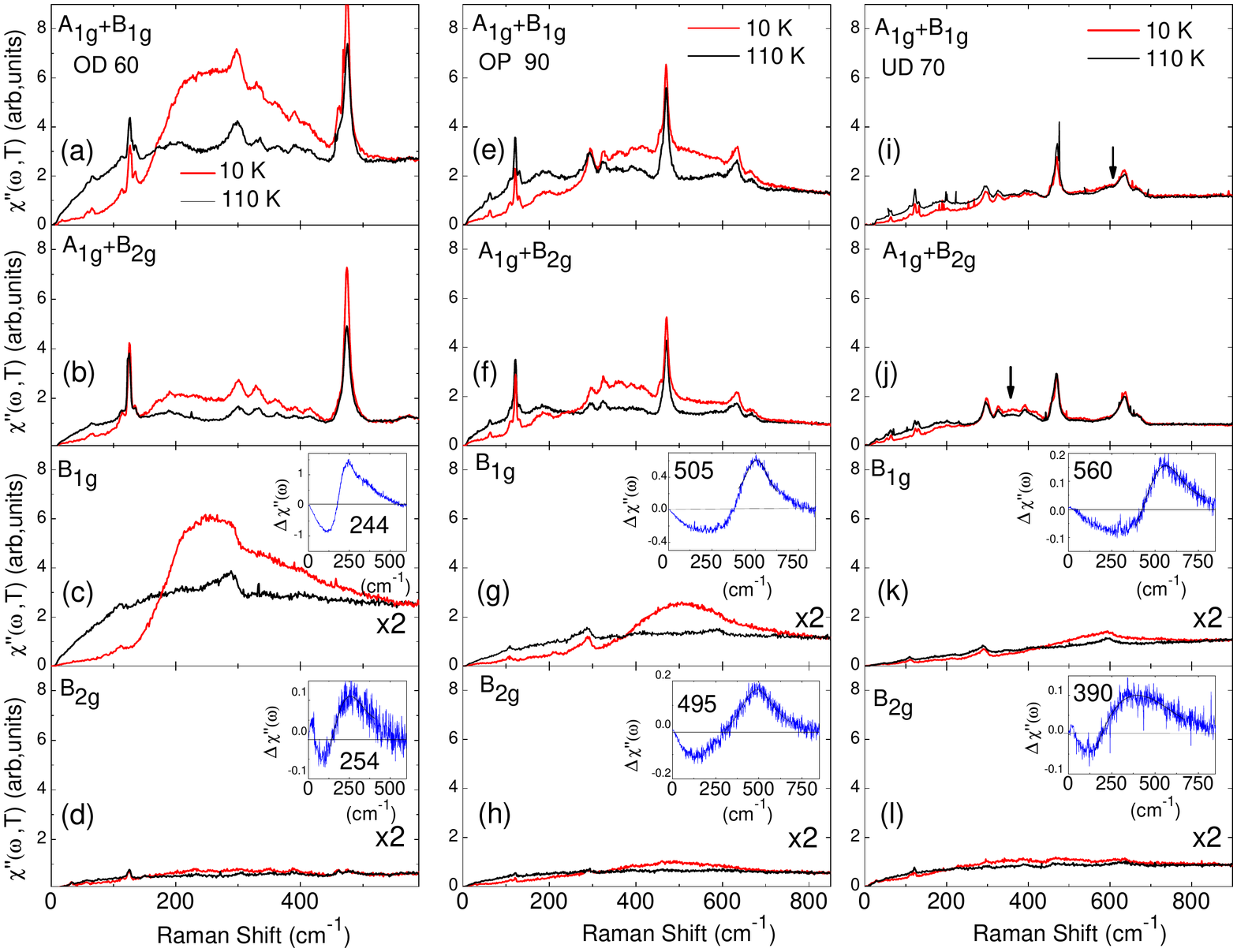}
\end{center}\vspace{-7mm}
\caption{(Color online) Raman responses $\chi^{\prime \prime}(\omega,T)$ in \AO+\BAN, \AO+\BN, \BAN and \BN symmetries in the normal and superconducting state of Bi-2212 single crystals for selected doping levels: p>0.22 (OD60), p= 0.16 (OP90) and p=0.11 (UD70). The insets show the Raman response at 10 K subtracted from the one at 110 K ($\Delta\chi^{\prime \prime}(\omega)$). This allows us to determined the energies of the \BAN and \BN electronic peaks.}
\label{fig1}
\end{figure}

The Raman responses in fig.1 (e-h) and (i-l) correspond respectively to the optimally doped crystal (OP) ($T_{c}= 90~K$), p=0.16 and the under-doped crystal (UD) ($T_{c}= 70~K$), p=0.11. Examining the \BAN and \BN Raman responses of the OP90 crystal (g,h), we can notice that the \BAN electronic contribution remains predominant compared to the \BN one although the \BAN Raman response is significantly reduced in intensity with respect to OD 60 compound. 
The energies of the \BAN and \BN superconducting peaks (see insets of panels (g-h)) are still close to each other. 
Interestingly the energy of the \AO peak is observable in \AO+\BN symmetry (panel (f)) 
and its energy is clearly distinct ($\approx350~cm^{-1}$) from those of \BAN and \BN peaks rather located around $500~cm^{-1}$. Concerning the UD 70 Bi-2212 single crystal, the \BAN and \BN superconducting peaks are clearly distinct in energy and respectively located at 560 $cm^{-1}$ and 390 $cm^{-1}$ (see insets of panels (k-l)). The weak fingerprints of these two peaks are indicated by black arrows in the \AO+\BAN and \AO+\BN spectra (panels (i-j)). On the other hand the \AO electronic peak is no more detectable in UD Bi-2212 single crystal. 

Importantly, we cannot easily extract the pure \AO electronic spectrum by subtracting the \BAN spectrum from the \AO+\BAN one (or the \BN spectrum from the \AO+\BN). There are two reasons for this: (i) the \BAN and \AO+\BAN spectra (or the \BN and \AO+\BN spectra) were obtained from two distinct crystal orientations (by rotating the crystal), this potentially introduces change in the Raman responses intensity; (ii) an additional half-wave plate for measuring the \BAN (or \BN) spectrum was used to keep the same polarization at the entrance of the spectrometer than the \AO+\BAN (or \AO+\BN) spectrum.  

In order to circumvent these difficulties and extract the pure \AO electronic contribution we made two distinct subtractions:\\
$\chi^{\prime \prime}_{\AOO}= \chi^{\prime \prime}_{\AO+\BAN}-\alpha\beta\chi^{\prime \prime}_{\BAN}$ (1)  \\
$\chi^{\prime \prime}_{\AOT}= \beta\chi^{\prime \prime}_{\AO+\BN}-\alpha\chi^{\prime \prime}_{\BN}$   (2) \\
$\alpha$ is the intensity correction factor corresponding to the half-wave plate absorption used in \BN and \BAN symmetries, $\alpha=1.2$ and $\beta$ is the intensity correction factor linked to the change of the crystal orientation between the \BAN and \AO+\BAN symmetries (or the \BN and \AO+\BN symmetries). 

In order to fix $\beta$ we defined two sums \cite{LeTaconPHD2006}:\\
$\chi^{\prime \prime}_{S1}=\chi^{\prime \prime}_{\AO+\BAN}+\alpha \chi^{\prime \prime}_{\BN}$ (2) and \\
$\chi^{\prime \prime}_{S2}=\chi^{\prime \prime}_{\AO+\BN}+\alpha \chi^{\prime \prime}_{\BAN}$ (3)

The $\chi^{\prime \prime}_{S1}$ and $\chi^{\prime \prime}_{S2}$ responses were successively obtained after a crystal rotation (see experimental procedure). The $\beta$ factor is then defined such as 
$\chi^{\prime \prime}_{S1}=\beta\chi^{\prime \prime}_{S2}$. The $\chi^{\prime \prime}_{S1}$ (solid line) and $\beta\chi^{\prime \prime}_{S2}$ (dashed line) responses in the normal and superconducting state are displayed in fig. 2 (c,f,i). They merge perfectly in the normal and superconducting states which allows us to estimate $\beta$ precisely for each doping level. $\beta$ is close to 1 with a variation less than 10 \%.

In fig.2 (a-b), (d-e) and (g-h) are displayed the pure \AO Raman spectra extracted in two distinct manners (\AOO and \AOT) in the normal and superconducting states for the OD60, OP90 and UD70 Bi-2212 single crystals. In the insets, the \AO electronic superconducting peak is revealed by subtracting the normal \AO contribution from the superconducting one. We find well defined \AO superconducting peaks consistent each other in the both cases of extraction which makes reliable our findings.  
Interestingly the energy of the \AOO superconducting peak (see inset) increases with under-doping ($\approx265~cm^{-1}$ to $\approx370~cm^{-1}$) while the intensity of the \AOO peak decreases before disappearing below p=0.12 (see insets of panels (a,d)).  

\begin{figure}[bp]
\begin{center}
\includegraphics[width=8.5cm,height=8.5cm]{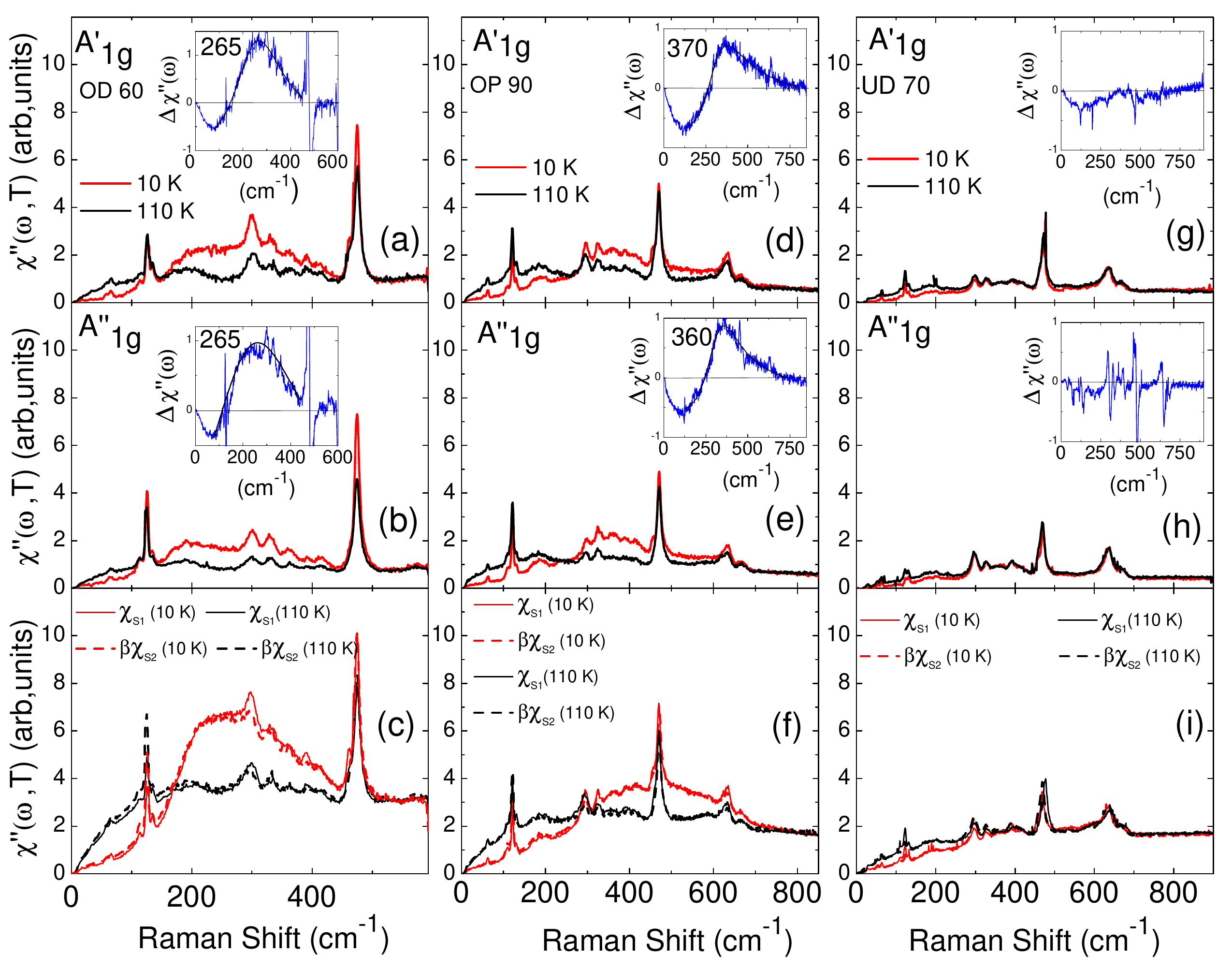}
\end{center}\vspace{-7mm}

\caption{(Color online) \AO extracted electronic Raman responses in the superconducting state (10 K) and normal state (110 K) for selected doping levels of Bi-2212 single crystals. 
Here are shown the distinct extractions of the \AO Raman response: \AOO and \AOT associated respectively with equations (1) and (2). In the insets are displayed the subtracted Raman responses at 10 K from those at 110 K.  The Raman responses in panels (c,f,i) show the sums $\chi^{\prime \prime}_{S1}$ and $\beta\chi^{\prime \prime}_{S2}$ related to the equations (1) and (2). $\beta$ was defined to make the sums merging.}

\label{fig2}
\end{figure}

In order to get a global view of the doping evolution of the three superconducting peaks detected in the \AO, \BAN and \BN symmetries, 
The energy of each superconducting peak as a function of doping is plotted in fig.3 (a). 

\begin{figure}[htp!]
\begin{center}
\includegraphics[width=7.5cm,height=8.5cm]{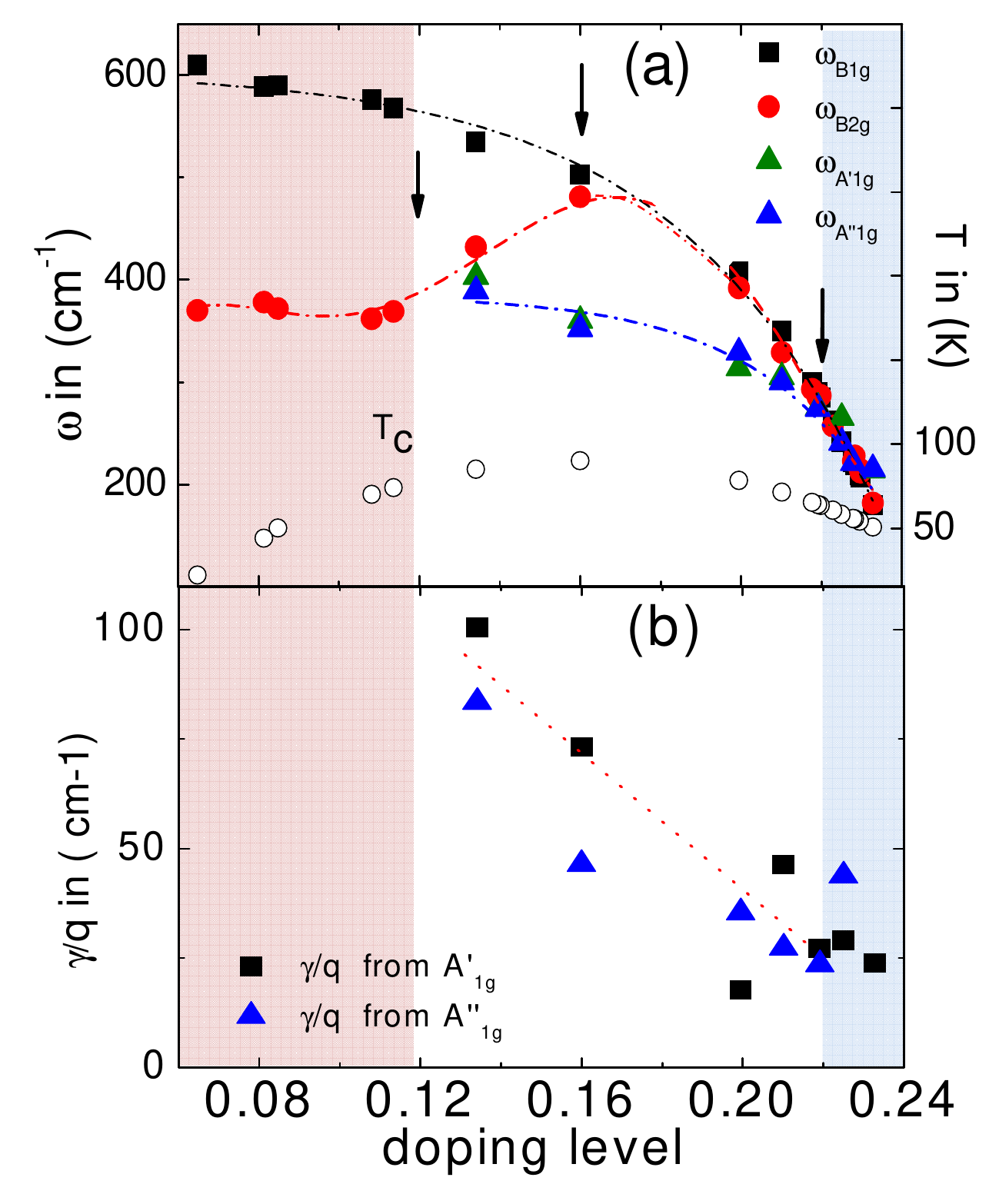}
\caption{(Color online)
(a) Doping evolutions of the \AO, \BAN and \BN energy scales in the superconducting state. The arrows pinpoint the three special doping levels: Above p=0.22 (blue/ grey) zone, all the energies scales merge, at p=0.16 the \BN scale is maximum and below p=0.12, the \BN and \BAN scales are almost constant in energy while the \AO one is no more detectable. Open circles describe the superconducting transition $T_{c}$ versus \textit{p} following the Presland-Tallon law \cite{Presland}; (b) asymmetric coefficient extracted from \AOO and \AOT superconducting peaks by a standard fit (see equation (4)).}
\label{fig3}
\end{center}\vspace{-7mm}
\end{figure}


Firstly, we find that the three energies merge above p=0.22. Below p=0.22, the \AO energy moves away from the \BAN and \BN scales. This is confirmed from the both extractions (\AOO or \AOT). The \AO scale is no more detected below p=0.12 while the \BAN and \BN scales are detected until low doping level (0.07). The \BAN energy scale monotonically increases with under-doping and seems to saturate below p= 0.12. 

The \BN scale is non monotonic and exhibits a maximum close to the optimal doping p=0.16. Above p=0.16 the \BN decreases and it becomes constant in energy below p=0.12. Our study of the \BN and \BAN energy scales at low and high doping levels is a refinement of earlier works 
on Bi-2212~\cite{Blumberg1997,Gasparov1998,Naeini1999,Opel2000,Sugai2000,Hewitt2002,Masui1,Blanc2010,Munnikes2011}. Here we show that the \BAN and \BN energy scales vary very slowly  below p=0.12 down to p=0.06 . 
The  saturation of the \BAN superconducting peak in energy was previously reported down to $p\approx0.11$ and interpreted as a pseudo resonance mode stemming from a strong fermionic self-energy due to the interaction with spin fluctuations~\cite{Chubukov1999}. 
Our extended work to low and high levels of doping reveals two particular ranges: (i) above p=0.22 (blue/grey) zone in fig 3 (a) where \AO,\BAN and \BN superconducting peaks merge and (ii) below p=0.12 (red/ grey) zone in fig.3(a) where the \BAN and \BN peaks are almost constant in energy. 

This is illustrated in figs.4 and 5. In fig.4 are displayed the \BAN, \BN and \AO Raman responses 
of two Bi-2212 single crystals in the superconducting and normal states for p$\geq0.22$. The subtracted Raman responses at 10 K from the one at 110 K show that for OD50 and OD63 compounds, the energies of the \BAN and \BN and \AO peaks coincide with each other (see panels (d-f) and (i-k)). 

\begin{figure}[htp!]
\begin{center}
\includegraphics[width=8.5cm,height=7cm]{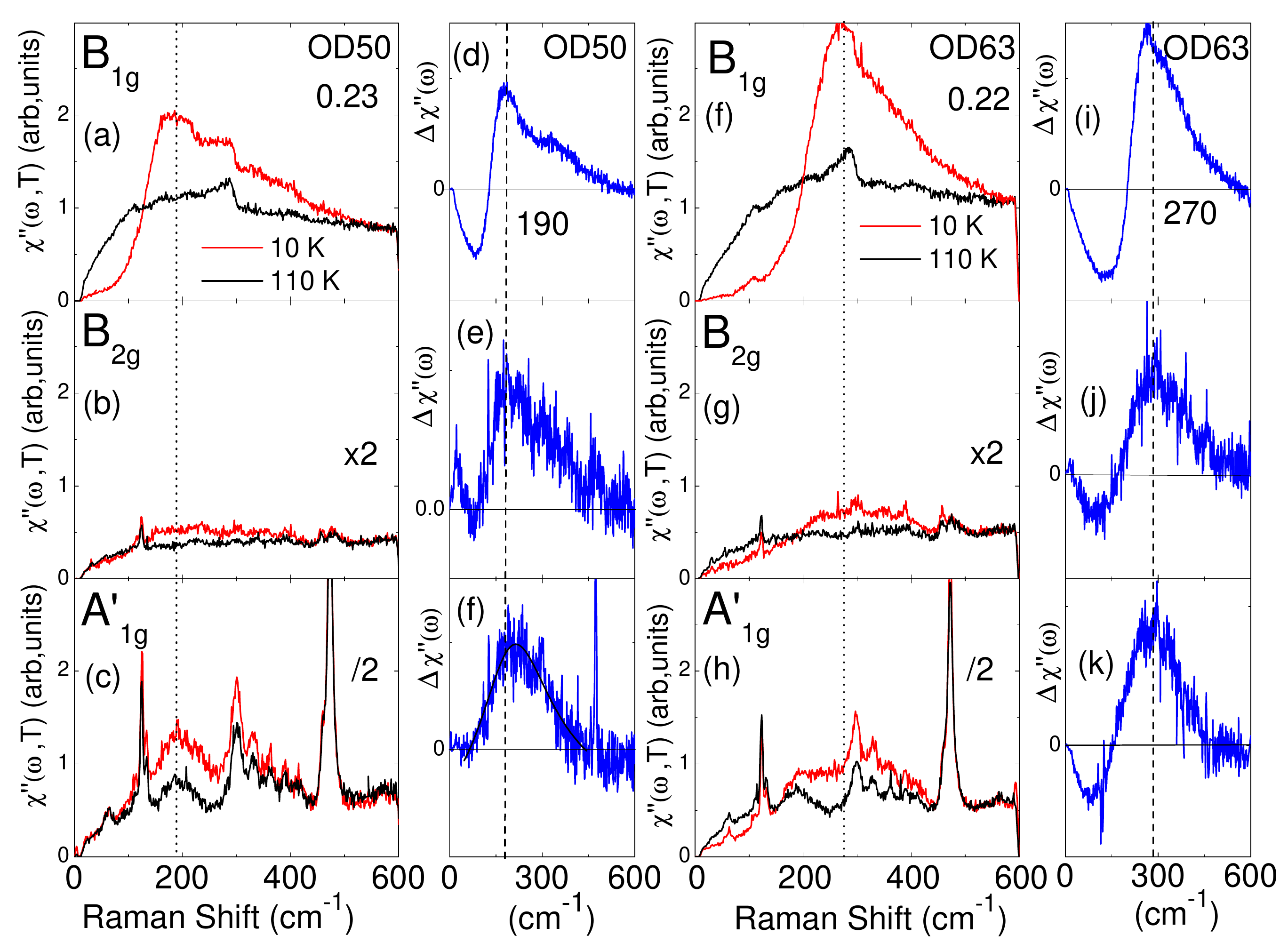}
\caption{(Color online)
\BAN and \BN and \AO Raman responses in the superconducting (red/grey) and normal (black) states of over-doped Bi-2212 single crystals. (a-c) for p=0.23, (g-i) for p=0.22. (d-f) and (j-l) correspond respectively to the subtracted Raman responses at 10 K from the one at 110 K for p=0.23 and p=0.22. The dashed line show that the locations in energy of the \BAN, \BN and \AO merge for each doping level above p=0.22.}
\label{fig4}
\end{center}\vspace{-7mm}
\end{figure}

On the other hand, in fig. 5 are reported selected \BAN and \BN Raman responses of under-doped Bi-2212 crystals (below p=0.12) in the superconducting and normal states. The Raman responses measured at 10 K subtracted from the one at 110 K (see insets of the panels (a-f)) show that the \BAN energy scale saturates and the \BN scale is almost constant in energy. 
\begin{figure}[htp!]
\begin{center}
\includegraphics[width=8.5cm,height=7cm]{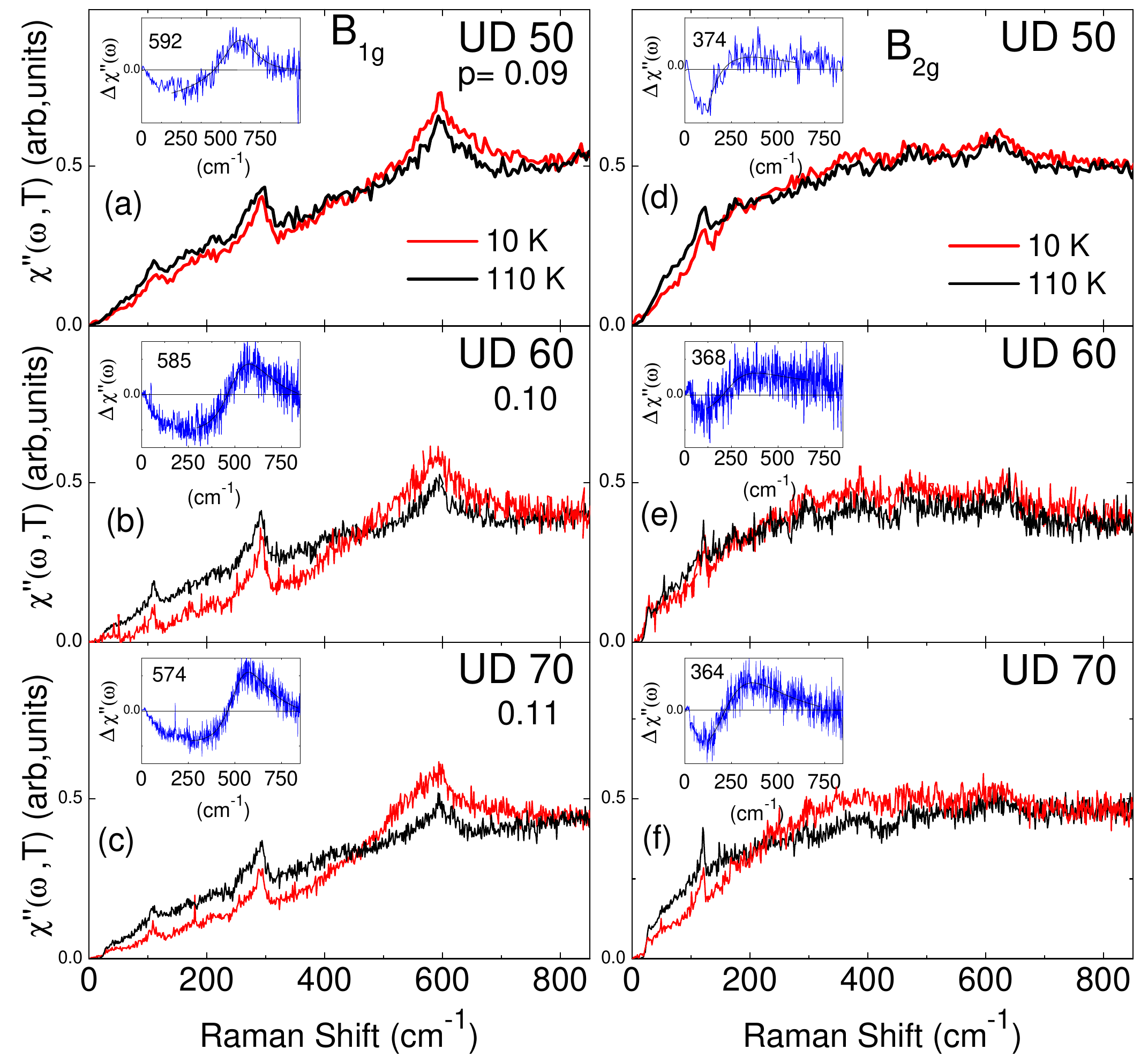}
\caption{(Color online)
\BAN  and \BN  Raman responses ($\chi^{\prime \prime} (\omega,T)$) in the superconducting (red/grey) and normal (black) states of selected underdoped Bi-2212 single crystals below p=0.12. 
The inset in each panel corresponds to the subtracted Raman response ($\Delta\chi^{\prime \prime}(\omega)$) obtained from measurements at 10 K and 110 K.  We can notice that both \BAN, \BN electronic peak are almost constant in energy below p=0.12. Note that the Raman spectra of UD 50 compound was obtained with a 600 g/mm diffraction grating instead of 1800 g/mm one in order to improve the signal/noise ratio and detect the \BAN and \BN superconducting peaks which become weaker in intensity with underdoping. As a consequence, the energy resolution of UD50 compound is lower than the other doping.}
\label{fig5}
\end{center}\vspace{-7mm}
\end{figure}


Three special doping levels are pinpointed in the evolutions of the \AO, \BAN and \BN scales (see black arrows in fig.3 (a)): (i) p=0.22 above which the three energy scales merge, (ii) p=0.12 below which the \AO scale disappears whereas the \BN and \BAN scales become constant in energy and finally (iii) $p=0.16$ where the \BN scale is peaked and reaches a maximum.  
We suspect that at least two of these particular doping levels arise from Fermi surface evolution. It is surprising to find that above p=0.22 the three energy scales merge. We rather expect for a d-wave superconducting gap that the \AO, \BAN and \BN scales have quite distinct energies. Considering a full hole-like cylindrical Fermi surface the energy ratios between the three scales should be \cite{Devereaux-RMP}: $\omega_{\BAN}/\omega_{\BN}\approx1.3$ and $\omega_{\BAN}/\omega_{\AO}\approx3$ instead of $\omega_{\BAN}/\omega_{\BN}\approx\omega_{\BAN}/\omega_{\AO}\approx1$. Such a discrepancy have also been reported in strongly overdoped Tl$_{2}$Ba$_{2}$CuO$_{6+\delta}$ (Tl-2201) and Y-123 cation-substituted compounds. This has been interpreted as a change of the d-wave gap symmetry into a mixing of a d-wave gap with a s-wave component \cite{Nishikawa2008,Masui2003}. However, the thermal conductivity measurements in strongly overdoped Tl-2201 showed that nodes are still there \cite{Proust2002}. Consequently, we rather believe in  
a strong alteration of the quasiparticles spectral weight in the antinodal Raman response which modifies the \AO,\BAN and \BN electronic peaks positions. This is indeed the case, in ARPES measurements on Tl-2201 in the superconducting state, a quasiparticles anisotropy reversal between the nodes and the antinodes was reported. At high doping level the low energy antinodal Bogoliubov quasiparticle spectral weight is strongly enhanced with respect to the nodal one \cite{Plate2005}.

In the specific case of Bi-2212 case, the doping level p=0.22 corresponds to the Lifshitz transition wherein, as a van Hove singularity crosses the chemical potential with underdoping, the electron-like anti-bonding Fermi surface at high doping level transforms into hole-like. A such a change in the Fermi surface topology considerably increases the antinodal quasiparticles spectral weight and can make the three energy scales merge above p=0.22.  Preliminaries calculations advocate for this scenario but more deeper theoretical investigations are still required and will be developed in a near future.  The Lifshitz transition in overdoped Bi-2212 has been first observed by ARPES in Bi-2212 ~\cite{Kaminski} and recently detected by the analysis of the integrated Raman intensity as a function of doping level in Bi-2212 ~\cite{benhabib15}. We found that p=0.22 is the starting point of the pseudogap as the doping decreases. The occurrence of a Lifshitz transition at the doping level for which the pseudogap collapses has also been reported in other cuprates ~\cite{Ino,Piriou}.

On the other hand, when approaching the doping level p=0.12, the Raman intensities of the \BAN and \AO peaks are drastically reduced in a such a way that the \AO peak is no more detectable in the Raman spectra (see fig.1 (a,e,i) and  fig.2 (a,d,g) ;(b,e,h) for the \BAN and \AO peaks respectively). The decrease of the \BAN peak intensity implies that one of the \AO peak is also altered because the \AO peak is considered as a bound state of the \BAN pairing peak \cite{Venturini2000,Montiel2015}. 
We interpret this significant drop of the peak intensity as a loss of coherent Bogoliubov quasiparticles spectral weight at the antinodes while the nodal region is protected in the underdoped side of the cuprate phase diagram \cite{Blanc2010,Blanc2009,Sacuto13}. 
This has been corroborated by (i) ARPES measurements \cite{Ding01,Kondo2009} and  by (ii) scanning tunneling spectroscopy (STS) which show that Bogoliubov quasiparticles occupy only a restricted region in the k-space around the nodes \cite{Kohsaka2008}. We also infer that the Fermi arcs detected in the normal state by ARPES \cite{Norman1998} extend into the superconducting state and correspond to a loss of Bogoliubov quasiparticles spectral weight at the antinodes. The antinodal quasiparticles are no more available for the superconducting state because they are involved in a distinct electronic order.
The loss of antinodal quasiparticles spectral weight starts below the pseudogap temperature $T^*$ and it is strongly accentuated  by the emergence of the charge ordering as the temperature decreases.

Charge ordering was first proposed from STS measurements in underdoped Bi-2212~\cite{McElroy2005,Wise2008,Parker2010} and then revealed by nuclear magnetic resonance \cite{Wu11,wu2015}, resonant X-ray scattering ~\cite{Ghiringhelli12,Chang12,daSilvaNeto2014} and recently confirmed by tunneling \cite{Fujita14}.
Interestingly, the temperature onset of the Fermi surface reconstruction defined as the temperature for which the Hall coefficient towards negative values in Y-123~\cite{LeBoeuf2007,Doiron-Leyraud13,Cyr2015} is also peaked at p=0.12. The observation of a sign change in the Hall coefficient at low temperatures hints that Fermi surface undergoes a reconstruction into pockets induced by some form of (short-range) charge ordering under high magnetic field \cite{Grisonnanche}. 

The loss of antinodal spectral weight in the electronic Raman spectra has a clear impact on the \BAN and \BN peak energies as showed in previous works \cite{Chen1997,Chubukov2008,LeBlanc2010,Blanc2009,Blanc2010,Sacuto13}. When quasiparticles spectral weight is mostly concentrated around the nodal regions, this pushes back the \BN peak to lower energy and makes larger the \BAN/\BN energy ratio with respect to this expected value close to 1.3. This can be seen in Fig.3 (a). Below p=0.12 the \BAN/\BN  ratio is $\approx1.6$ higher than the one expected. The saturation of the pairing gap energy at the antinodes (\BAN scale) can then be interpreted as a competition between superconductivity and charge ordering which prevents an increase of $T_c$ in agreement with recent investigations \cite{Cyr2015,Cyr2015b}.

Finally p=0.16 corresponds to the doping level for which the superconducting transition $T_{c}$ is maximum. This could be the best compromise to enhance superconductivity in a complex medium where several electronic phases compete.

Interestingly the line shape of the \AO superconducting peak becomes more asymmetric with under doping. See for instance its change in the inset of fig.2(d) (OP90) in comparison with the line shape of fig.2(a) (OD60). We can estimate the line shape asymmetry as a function of doping by fitting the subtracted Raman responses $\Delta\chi^{\prime \prime}_{\AOO, \AOT}(\omega) =\chi^{\prime \prime}_{\AOO,\AOT} (\omega,10K)- \chi^{\prime \prime}_{\AOO,\AOT} (\omega,110K)$  
by a standard line shape equation : \\  

$\Delta\chi^{\prime \prime}_{\AOO,\AOT}(\omega)= \frac{A}{q^2\gamma}\frac{((\omega-\omega_0)/\gamma+q)^2}{1+ ((\omega-\omega_0)/\gamma)^2}$      (4)\\

where $A$ is a renormalization factor,  $q$ the asymmetric coefficient. $q$ tends to infinity for a Lorentzian shape and to a finite value otherwise. $\gamma$ is the full width at half maximum and $\omega_0$ is defined such as the \AO peak takes a maximum value at $\omega_{m}= \omega_0 + \frac{\gamma}{q}$. We chose to report  the $\frac{\gamma}{q}$ ratio as the strength of the asymmetric line shape.

The $\frac{\gamma}{q}$ ratio versus doping is plotted in fig. 3 (b). It has been calculated from both $\Delta\chi^{\prime \prime}_{\AOO}(\omega)$ and  $\Delta\chi^{\prime \prime}_{\AOT}(\omega)$ subtracted responses. We find in both cases an increasing of the asymmetric line shape with under-doping. We suspect this asymmetry comes from the additional contribution $2\Delta$ pairing peak to the \AO peak as already reported in our previous study \cite{LeTacon05b}. The asymmetric line shape is then enhanced by the increase of the distance in energy between the \AO peak and the $2\Delta$  pairing peak as the doping level is reduced (see fig.3 (a)).  

In conclusion, we have succeeded to extract from electronic Raman scattering study in a reliable manner, the \AO, \BAN and \BN Raman responses. We find three electronic peaks distinct in energy related to the three symmetries in the superconducting state of Bi-2212. We report a finely tuned doping evolution of these three electronic peaks by oxygen insertion only into the Bi-2212 structure without cationic substitution. The \BAN and \AO increase monotonically as the doping level is reduced whereas the \BN exhibits a non monotonic behavior with a maximum near the optimal doping level. We identify three special doping levels. Above p=0.22 all the peaks merge in energy (see blue zone in fig.3(a)). Below, p=0.12 the \AO is no more detected whereas the \BAN and \BN scales exhibit two "`plateau"' at distinct energies (see red zone in fig.3 (a)). Between p=0.22 and p=0.12,  the \AO peak and \BAN scales increases monotonically while the \BN scale is non monotonic and exhibits a maximum in energy at p=0.16. 
We suspect that at least the p=0.22 and 0.12 doping levels are directly connected to the doping evolution of the antinodal Bogoliubov quasiparticles spectral weight at low energy. p=0.22 corresponds to the doping level for which a Lifshitz transition occurs \cite{benhabib15,Kaminski} and the antinodal quasiparticles spectral weight is strongly increased. On the other hand, p=0.12 corresponds to the doping level for which the charge ordering is well settled \cite{Ghiringhelli12,Chang12,Wu11,wu2015,McElroy2005,Wise2008,Parker2010} and the antinodal quasiparticles spectral weight is strongly reduced. Finally, p=0.16 corresponds to the maximum of $T_{c}$ where probably the Fermi surface is still disturbed. The \BAN, \BN and \AO peaks all disappear above $T_{c}$. The \BAN and \BN peaks are two pieces of the $2\Delta$ pairing gap depending on the the part of the Fermi surface probed and the \AO peak is a collective mode related to the $2\Delta$ pairing gap, probably a bound state located below $2\Delta$ threshold of the particle-hole continuum~\cite{Montiel2015}.  
Remarkably, these three superconducting energy scales although partially detected in other cuprates such as HgBa$_2$CuO$_{6+\delta}$, (Hg-1201), Y-123 and Tl-1201 are an universal feature to all the cuprates and reveal competing orders inside the superconducting dome. 

We are grateful to C. P\'epin, X. Montiel, M. Civelli, I. Paul, Ph. Bourges, Y. Sidis, M. H. Julien, M. LeTacon and A. Georges. Correspondences and requests for materials should be addressed to A.S. (alain.sacuto@univ-paris-diderot.fr)

\bibliography{cuprates}

\end{document}